\documentclass[11pt, a4paper]{article}
\usepackage{float}
\usepackage[utf8]{inputenc}
\usepackage{geometry}
\usepackage{cite}
\usepackage{amsmath,amssymb,amsfonts}
\usepackage{graphicx}
\usepackage{textcomp}
\usepackage{xcolor}
\usepackage{hyperref}
\usepackage{authblk} 
\usepackage{orcidlink}
\usepackage{booktabs} 
\usepackage{multirow} 
\usepackage{array}    

\usepackage{algorithm}
\usepackage{algorithmic}

\providecommand{\keywords}[1]
{
  \small	
  \textbf{\textit{Keywords---}} #1
}

\begin{document}

\title{Enhancing Analogy-Based Software Effort Estimation with Firefly Algorithm Optimization}


\author[1]{Tarun Chintada\,\orcidlink{0009-0002-1528-0714}\thanks{Equal contribution. Work performed while at Jawaharlal Nehru Technological University - Gurajada Vizianagaram. Current address: Department of Computer Science and Engineering, Indian Institute of Technology Patna. \protect\\ \texttt{tarunchintada2004@gmail.com}}}

\author[1]{Uday Kiran Cheera\thanks{Equal contribution. \protect\\ \texttt{cheeraudaykiran@gmail.com}}}

\affil[1]{Department of Information Technology, Jawaharlal Nehru Technological University – Gurajada Vizianagaram, Vizianagaram, India}

\date{} 

\maketitle

\begin{abstract}
Analogy-Based Estimation (ABE) is a popular method for non-algorithmic estimation due to its simplicity and effectiveness. The Analogy-Based Estimation (ABE) model was proposed by researchers, however, no optimal approach for reliable estimation was developed. Achieving high accuracy in the ABE might be challenging for new software projects that differ from previous initiatives. This study (conducted in June 2024) proposes a Firefly Algorithm-guided Analogy-Based Estimation (FAABE) model that combines FA with ABE to improve estimation accuracy. The FAABE model was tested on five publicly accessible datasets: Cocomo81, Desharnais, China, Albrecht, Kemerer and Maxwell. To improve prediction efficiency, feature selection was used. The results were measured using a variety of evaluation metrics; various error measures include MMRE, MAE, MSE, and RMSE. Compared to conventional models, the experimental results show notable increases in prediction precision, demonstrating the efficacy of the Firefly-Analogy ensemble.
\end{abstract}


\section{Introduction}
Estimating software development effort is a vital and often unpredictable aspect of project management. According to Fadhil et al.\cite{b5}, predicting project expenses in the first stage is a significant step because it is crucial to calculate the resources and budget required to deliver the project. It is difficult to manage a software project without exact estimates and planning. With time, effort estimation models have grown in sophistication, presenting more reliable forecasts than previous ones, but they are still unable to provide significantly more accurate estimates \cite{b1}\cite{b2}. In 2017, a survey by the Project Management Institute (PMI) found that 69\% of software projects met their goals and business intent, while 43\% were not completed within budget, 48\% were delivered late, and 15\% failed due to inaccurate effort estimation \cite{b8}.

In recent years, machine learning (ML)-based algorithms have gained popularity in Software Development Effort Estimation (SDEE) research. Some researchers \cite{b10,b11,b12} consider Machine learning (ML) approaches are regarded as one of the three primary estimation techniques, with expert judgment and algorithmic models constituting the other two. Boehm and Sullivan \cite{b13} classified learning-oriented strategies as one of six software effort estimation techniques. Similarly, Zhang and Tsai \cite{b14} examined the implementation of various ML methods in the SDEE domain, including case-based reasoning, decision trees, artificial neural networks, and genetic algorithms. Wen's (2012) study \cite{b15} indicates that most ML models deliver satisfactory accuracy, typically achieving arithmetic mean MMRE values between 35\% and 55\%, Pred(25) figures ranging from 45\% to 75\%, and MdMRE values from 30\% to 50\%. Specifically, the mean MMRE for ANN, CBR, and DT is approximately 35\%, 50\%, and 55\%, respectively.

Analogy-Based Estimation (ABE), a type of CBR, has lately emerged as a promising strategy, with some studies indicating comparable accuracy to algorithmic methods, and it may be easier to comprehend and apply. ABE estimates project costs by identifying the most relevant project from a pool of historical projects (\cite{b10}). Developing similar software projects requires comparable work, making it the most extensively used technique in research and practice \cite{b3}. Wen et al. \cite{b15} found that analogy-based reasoning is the most commonly utilized machine learning technique for software effort estimation. Ease of use may be an essential component in the effective adoption of estimate methods within industry. Therefore, analogy-based estimation warrants additional scrutiny\cite{b7}.

Various approaches, including Neural Networks, Fuzzy Logic, Particle Swarm Optimization (PSO), Nearest Neighbors, and Genetic Algorithms (GA), have been applied to enhance the accuracy of software development effort estimation \cite{b18}, \cite{b11,b12,b13,b14,b15,b16,b17}. These optimization techniques focus on refining feature selection or attribute weighting within the similarity function used in analogy-based estimation (ABE). Many of these methods draw inspiration from nature: PSO mimics the coordinated movement of birds and fish, Ant Colony Optimization is based on the foraging behavior of ants, the Artificial Bee Colony algorithm replicates the search patterns of bees, and the Firefly Algorithm (FA) takes its cues from the mating behavior of fireflies.

FA excels at optimizing analogy selection in ABE because to its ability to handle complicated, multidimensional issues with nonlinear characteristics \cite{b30}. Research suggests that FA outperforms other optimization strategies such as PSO, GA, and ABC \cite{b28,b29} because to its global search mechanism that identifies the strongest parallels. As a result, inspired by the preceding, this work seeks to combine FA and ABE to improve software development effort estimation.

The introductory portion presents the background and associated works, while ABE and FA's linked works are described in their separate sections. The remainder of the paper is divided into two sections: II and III, which discuss the ABE and Firefly Algorithm (FA), respectively. Section IV contains details about the suggested model. The findings are presented in Section V, followed by the conclusion in Section VI. 

\section{ANALOGY-BASED ESTIMATION (ABE)}
Shepperd and Schofield introduced ABE, a non-algorithmic estimate approach \cite{b10}. This is a kind of Case-Based Reasoning (CBR). Case-Based Reasoning (CBR) assumes similar software projects have comparable costs. When working on a global software project, it's important to identify relevant and independent attributes. Next, compare the proposed project to all other projects in the database. The estimated effort for the new project is based on known values from previous initiatives. There are usually four elements to ABE,
\begin{enumerate}
    \item Historical Projects 
    \item Identification of project characteristics
    \item Utilization of defined similarity functions
    \item Project effort estimation
\end{enumerate}
Each of them can be defined as:
\begin{enumerate}
    \item Gathering information from previous projects to build a comprehensive historical database.
    \item Extracting the pertinent features of a new project that mirror those documented in the historical dataset.
    \item Applying standard similarity measures, such as Euclidean or Manhattan distances, to identify projects that closely resemble the new one based on shared attributes.
    \item Estimating the required effort for the new project by utilizing statistical functions, such as the mean and median.
\end{enumerate}

\subsection{SIMILARITY FUNCTION}
In analogy-based estimation (ABE), assessing the similarity between projects is crucial. This similarity function measures how alike two projects are \cite{b10}. Two primary similarity metrics are used: Euclidean Similarity (ES) and Manhattan Similarity (MS). The Euclidean Similarity is defined in Equation (1).

\begin{equation}
\text{Sim}(p, p') = \frac{1}{\sqrt{\sum_{i=1}^{k} w_i \cdot \text{Dis}(a_i, a_i') + \delta}} \quad \delta = 0.0001
\end{equation}

Here, the distance function is specified as follows:

\begin{equation}
\text{Dis}(a_i, a_i') =
\begin{cases}
|a_i - a_i'|, & \text{if } a_i \text{ and } a_i' \text{ are ordinal or numerical}, \\
0, & \text{if } a_i \text{ and } a_i' \text{ are nominal and } a_i = a_i', \\
1, & \text{if } a_i \text{ and } a_i' \text{ are nominal and } a_i \neq a_i'.
\end{cases}
\end{equation}

In this formulation:
\begin{itemize}
    \item $p$ and $p'$ represent the two projects being compared.
    \item $w_i$ is the weight attributed to each feature, with values ranging from 0 to 1.
    \item $\delta$ is a small constant (0.0001) to prevent division by zero.
    \item $a_i$ and $a_i'$ denote the individual attributes of the projects.
    \item $k$ indicates the total number of features considered.
\end{itemize}

Although the Manhattan and Euclidean similarity functions share similarities, the Manhattan Similarity (MS) calculates the absolute differences between features directly. The Manhattan Similarity is expressed in Equation (3):

\begin{equation}
\text{Sim}(p, p') = \frac{1}{\sum_{i=1}^{k} w_i \cdot \text{Dis}(a_i, a_i') + \delta} \quad \delta = 0.0001
\end{equation}

with the distance function defined as:

\begin{equation}
\text{Dis}(a_i, a_i') =
\begin{cases}
|a_i - a_i'|, & \text{if } a_i \text{ and } a_i' \text{ are ordinal or numeric}, \\
0, & \text{if } a_i \text{ and } a_i' \text{ are nominal and } a_i = a_i', \\
1, & \text{if } a_i \text{ and } a_i' \text{ are nominal and } a_i \neq a_i'.
\end{cases}
\end{equation}

\begin{itemize}
    \item $p$ and $p'$ denote the projects being compared.
    \item $w_i$ represents the weight for each attribute, with a range from 0 to 1.
    \item $\delta$ ensures that the denominator remains non-zero.
    \item $a_i$ and $a_i'$ refer to the attributes or features of the projects.
    \item $k$ is the number of features considered.
\end{itemize}

\subsection{SOLUTION FUNCTION}
Solution functions are essential for estimating effort based on historical project data. Techniques such as closest analogy, inverse weighted mean (IWM), mean, and median enable a comparison of the projected work with that of comparable projects \cite{b32}. Notably, the IWM method differentiates itself by assigning individual weights to each similar project, so that each one's contribution to the final estimate reflects its degree of similarity to the new project. This method of weighted aggregation emphasizes the relevance of each analogous project in forming the overall effort estimate. The computation of the inverse weighted average is demonstrated in equation (5):

\begin{equation}
    C_P = \sum_{k=1}^{S} \frac{\text{Sim}(p, p_k)}{\sum_{i=1}^{n} \text{Sim}(p, p_i)} \, C_{p_k}
\end{equation}

\begin{itemize}
    \item $p$: The new project for which the effort estimation is performed.
    \item $p_k$: The $k$-th project identified as being similar to the new project.
    \item $C_{p_k}$: The effort or cost associated with the $k$-th similar project.
    \item $S$: The total number of similar projects considered in the estimation process.
\end{itemize}

Xin-She Yang created the firefly algorithm (FA) in late 2007 and published it in 2008 \cite{b33,b34}. Since then, its literature has grown significantly with varied applications. Firefly behavior and flashing patterns served as the inspiration for FA. First, let's describe the flashing behavior of tropical fireflies.\cite{b27}

\subsection{Firefly Behaviour}
Fireflies illuminate the summer skies in tropical and temperate locations. Approximately 2,000 species of firefly emit brief, repetitive flashes. The flash patterns vary between species. Bioluminescence causes the flashes, although the exact purpose of this communication system is unknown. These flashes have two purposes: communicating with mating partners and attracting possible prey\cite{b21}. Furthermore, flashing may signal to potential predators that the firefly has an undesirable taste. The alarm system features rhythmic flashes, flash frequency, and flash interval. The signaling mechanism connecting the sexes includes rhythmic flashes, flash frequency, and flash interval \cite{b21}. Female fireflies react to the distinct flash patterns of males of the same species.In some species, such as Photuris, female fireflies eavesdrop on bioluminescent courtship signals and replicate other species' mating flash patterns, enticing and preying on male fireflies who mistake their flashes for appropriate partners. Some tropical fireflies demonstrate self-organizing behavior by synchronizing their flashes. We know that light intensity at a given distance from the light source follows the inverse-square law, as seen in Equation(6). 
\begin{equation}
    I \propto \frac{1}{r^2}
\end{equation}
Furthermore, air absorbs light, making it weaker as distance increases. Fireflies may communicate from a distance of several hundred meters at night due to a combination of visibility variables.

\subsection{Firefly-Inspired Algorithm}
To streamline the explanation of the standard Firefly Algorithm (FA), we employ three simplified principles:

\begin{itemize}
    \item All fireflies are treated as unisex, so any firefly can be attracted to any other, independent of gender.
    \item The attractiveness of a firefly is directly tied to its brightness. When two fireflies flash, the one with lower brightness moves toward the one emitting a stronger light. Both brightness and attractiveness diminish with increasing distance. In the absence of a brighter firefly, a firefly will choose a random direction.
    \item The brightness of each firefly is defined by the landscape of the objective function.
\end{itemize}

Equation (6) describes a firefly at position \(P' \) and brightness \(I \). \begin{equation}
    ( I(P') \propto f(x) )
\end{equation} The attraction \(\beta \) is proportional to the flies and connected with the distance \(D_{i,j} \) between firefly \(i \) and \(j \).

Equation (8) shows the inverse square of intensity \(I(r) \), with \(I_0 \) representing the intensity of light from the source.
\begin{equation}
    I(r) = I_0 e^{-\gamma r^2}
\end{equation}

Equation (9) calculates intensity assuming an absorption factor of the environment \(\gamma \), where \(I_0 \) represents the initial intensity.
\begin{equation}
    I(r) = \frac{I_0}{1 + \gamma r^2}
\end{equation}
Equation (10) represents the distance between fireflies at positions \(p_i \) and \(p_j \). \(p_{j,k} \) represents the \(k^{th} \) component of the spatial coordinate \(p_i \).
\begin{equation}
    D_{i,j} = \|p_i - p_j\| = \sqrt{\sum_{k=1}^d (p_{i,k} - p_{j,k})^2}
\end{equation}

A firefly \(i \) attracts a brighter one \(j \), as seen in Eq. (11). The attraction can be represented by \(\beta e^{-\gamma r_{i,j}^2} (p_i - p_j) \), and \(\alpha \left[\text{rand} - \frac{1}{2} \right] \) denotes the randomness based on the randomization parameter. \( \alpha \).
\begin{equation}
    p_i = p_i + \beta e^{-\gamma r_{i,j}^2} (p_i - p_j) + \alpha \left[ \text{rand} - \frac{1}{2} \right]
\end{equation}

Furthermore, \(\gamma \) controls attraction changes, which influence the behavior and convergence speed of FA.

\begin{figure}[h]
    \centering
    \includegraphics[width=0.7\textwidth, height = 300px]{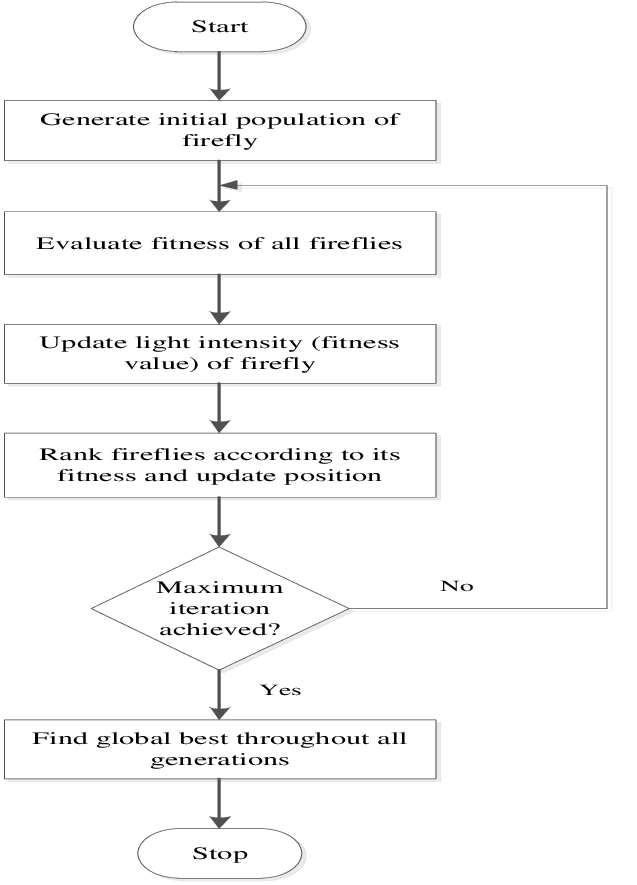} 
    \caption{The workflow of firefly algorithm.}
    \label{fig:workflow}
\end{figure}

\section{RELATED WORKS}

Project management relies heavily on software effort estimation, which has a direct impact on scheduling, resource allocation, and planning. A variety of techniques have been investigated to increase estimation accuracy, ranging from sophisticated machine learning and heuristic optimization to conventional statistical models. Because it uses historical project data to enable estimation based on parallels with prior projects, Analogy-Based Estimation (ABE) stands out among these. This method makes use of the realization that effort estimates can be guided by comparable project attributes. However, irrelevant or weakly correlated features might skew similarity calculations, resulting in erroneous effort estimations, and making ABE susceptible to input noise.

We looked into feature selection techniques to improve the ABE input space and lessen the influence of noise in order to overcome this constraint. In order to preserve features that were highly relevant to the target variable, we specifically used Pearson correlation with a threshold of 0.5. Despite the fact that this feature selection improved ABE similarity evaluations, accuracy variations among datasets indicated the need for additional improvement. In order to make sure that chosen features more correctly reflected project similarities, we investigated weight optimization strategies.

In addition to feature selection, we investigated a number of machine learning models and metaheuristic optimization strategies, such as support vector machines, neural networks, and ensemble approaches. Combining models can increase prediction accuracy, according to studies like Huang et al. (2004) on Extreme Learning Machines and other ensemble techniques. However, these methods can be unfeasible for project estimating in the real world due to their high processing demands and complexity. Furthermore, the unique problem of weight optimization in similarity-based techniques like ABE is not addressed by these models.

We used the Firefly Algorithm (FA) to optimize similarity weights in ABE in light of these findings. The FA is perfect for fine-tuning weight allocations within ABE because it effectively manages non-linear optimization and stays clear of local optima, which is modeled after the behavior of fireflies in the wild. By better-aligning characteristics and estimation targets, the resulting FAABE (Firefly Algorithm-Analogy-Based Estimation) model lowers prediction error and boosts reliability.

Five reputable datasets were chosen for assessment: Desharnais, COCOMO81, China, Albrecht, and Maxwell. These datasets were selected not only because they are widely available and well-liked, but also because of their representative and varied features, which enable a thorough examination of model performance across a range of project kinds. In particular:
\begin{table}[htbp]
\caption{Descriptive Statistics of the Datasets}
\label{tab:stats}
\centering
\renewcommand{\arraystretch}{1.2}
\begin{tabular}{l c c r r r}
\toprule
 &  &  & \multicolumn{3}{c}{\textbf{Effort Statistics}} \\
\cmidrule(lr){4-6}
\textbf{Dataset} & \textbf{Projects} & \textbf{Features} & \textbf{Min} & \textbf{Max} & \textbf{Median} \\
\midrule
\textbf{COCOMO81} & 64 & 16 & 5.9 & 11,400.0 & 102.0 \\
\textbf{Desharnais} & 81 & 12 & 546.0 & 23,940.0 & 3,647.0 \\
\textbf{China} & 499 & 14 & 26.0 & 54,620.0 & 1,829.0 \\
\textbf{Albrecht} & 24 & 8 & 0.5 & 105.2 & 11.5 \\
\textbf{Kemerer} & 15 & 7 & 23.2 & 1,107.3 & 130.3 \\
\textbf{Maxwell} & 62 & 27 & 583.0 & 63,694.0 & 5,189.5 \\
\bottomrule
\end{tabular}
\end{table}

We were able to assess the FAABE model across a range of project features and domains thanks to these datasets, which were selected for their representative nature and extensive use in software engineering research. Across all five datasets, our tests showed that FAABE continuously performed better than other machine learning models and conventional ABE. The model provides a workable, precise, and reliable solution for software effort estimation across a range of real-world settings by combining feature selection and FA-based weight optimization.

\section{Firefly Algorithm guided Analogy Based Estimation (FAABE)}
The recommended strategy combines the FA and ABE models to improve estimation accuracy. The FA has two essential properties that allow it to reduce the ambiguity and complexity of software project qualities: adaptation and flexibility. In essence, the FA's primary goal is to determine which feature weights are best suited for the similarity function. To enhance the performance of the ABE, weights are assigned for parameter tuning. While Algorithm 1 displays the FABE pseudo-code, Fig. 2 depicts the system designs for the training and testing of the suggested method.

\subsection{Training}
The design of the FABE approach's training phase is shown in Fig. 1. During the training phase, past project data is used to forecast the training dataset's efforts.
At this point, the model modifies the feature weights according to the Analogy-based Estimation similarity function's FA. The development effort is the dependent feature; all other elements are regarded as independent. All available dataset projects are separated into basic, train, and test subsets during the training phase. Throughout the training phase, models are developed using both basic and training subsets. The basic and test subsets are used to evaluate the model during the testing phase. Testing projects are compared to basic projects to evaluate their performance, and training projects are compared to basic projects to determine the necessary weight. 

\section*{Firefly-Based Estimation Algorithm}

\begin{algorithm}[H]
\caption{ABE with Firefly Optimization (FAABE)}
\label{algo:FAABE}
\begin{algorithmic}[1]
\REQUIRE Historical dataset of software projects (\(D\)), new project features (\(P\)), similarity function (e.g., Euclidean or Manhattan), solution function (e.g., weighted mean, mean, or median), Firefly Algorithm parameters: population size (\(N\)), light absorption coefficient (\(\gamma\)), randomness (\(\alpha\)), maximum iterations (\(T\))
\ENSURE Estimated project effort (\(E\))

\STATE \textbf{Step 1: Data Preprocessing} 
\STATE Collect and preprocess the dataset (\(D\)): normalize, clean, and standardize features.
\STATE Extract relevant features (\(F\)) using feature selection (e.g., Pearson correlation threshold of 0.5).

\STATE \textbf{Step 2: Similarity Computation}
\FOR{each project \(p' \in D\)}
    \STATE Compute similarity between \(P\) and \(p'\):
    \[
    Sim(P, p') = \frac{1}{\sqrt{\sum_{i=1}^{n} w_i \cdot Dis(f_i, f_i') + \delta}}
    \]
    \STATE Where \(Dis(f_i, f_i')\) accounts for differences (numerical or nominal).
\ENDFOR

\STATE \textbf{Step 3: Firefly Algorithm Initialization} 
\STATE Generate \(N\) fireflies, each representing feature weights (\(w_i\)).
\STATE Compute the initial brightness (\(I\)) of each firefly using the inverse weighted mean (IWM):
\[
I = \sum_{k=1}^{K} \frac{Sim(P, p_k)}{\sum_{i=1}^{n} Sim(P, p_i)} \cdot C_{p_k}
\]
\STATE Where \(C_{p_k}\) is the effort value of the \(k\)-th project.

\STATE \textbf{Step 4: Firefly Optimization}
\FOR{\(t = 1\) to \(T\)}
    \FOR{each firefly \(i\)} 
        \FOR{each firefly \(j\)}
            \STATE Compute distance between \(i\) and \(j\):
            \[
            R_{i,j} = \sqrt{\sum_{k=1}^{d} (x_{i,k} - x_{j,k})^2}
            \]
            \STATE Update position of firefly \(i\):
            \[
            x_i = x_i + \beta e^{-\gamma R_{i,j}^2} (x_j - x_i) + \alpha \left[\text{rand} - \frac{1}{2}\right]
            \]
        \ENDFOR
        \STATE Recompute brightness for all fireflies.
    \ENDFOR
\ENDFOR

\STATE \textbf{Step 5: Effort Estimation}
\STATE Select the brightest firefly (optimal weights \(w^*\)).
\STATE Recompute similarity and estimate effort (\(E\)) using the solution function with \(w^*\).

\RETURN \(E\)
\end{algorithmic}
\end{algorithm}

\begin{figure}[h]
    \centering
    \includegraphics[width=0.7\textwidth, height = 330px]{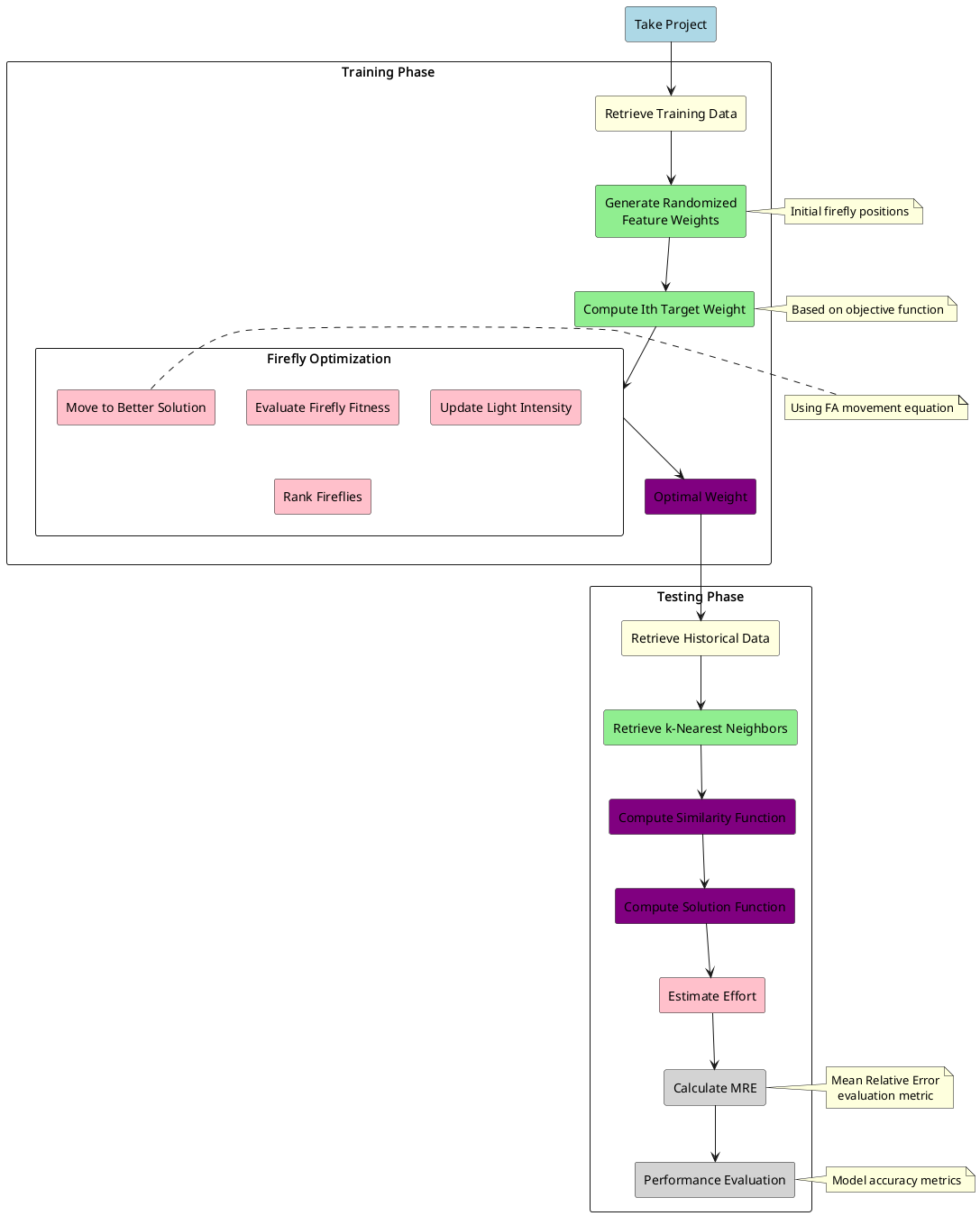}
    \caption{Proposed Methodology for FAABE Algorithm }
    \label{fig:Methodology}
\end{figure}

\subsection{Testing}
Phase of Evaluation  
The Testing Phase assesses the accuracy of the model by using both familiar and unfamiliar project data. It starts by leveraging the optimal feature weights established during the Training Phase to collect historical data and determine the kk-nearest neighbors of a particular test project. By properly prioritizing features that significantly influence effort estimates, these weights improve the similarity function.

For every test project:
\begin{itemize}
\item Acquire historical project data and utilize the similarity function to evaluate how closely a project aligns with others in the dataset.
\item  To estimate the effort, identify the kk most comparable projects, and pass them through a specified solution function (such as weighted mean, mean, or median).
\item Determine the mean relative error (MRE) of the estimated effort by comparing it to the actual effort of the test project.
\end{itemize}
This process is done for each test project in the dataset. Model performance metrics, such as accuracy and overall MRE, are calculated after all projects have been evaluated. To improve the accuracy of effort projections, the Testing Phase incorporates feature weights obtained during the Training Phase, as shown in Fig. 2.

Dataset used for the ABE model is partitioned into three sections, with one-third designated for testing and the remaining two-thirds for training and weight optimization. The feature weights established by the Firefly Algorithm are continually modified throughout the training process, resulting in fluctuating weight values across different execution runs. Rather than signifying static feature significance, these weights represent the ideal combination required for precise effort estimation.  
This flexible weighting strategy guarantees that the ABE model can adjust to various project datasets, ultimately enhancing accuracy in effort estimation tasks.

\section{EXPERIMENTAL RESULTS}

The mean of the individual outcome of every instance in the dataset is represented by the performance metrics that were reported:MMRE, MAE, MSE and RMSE.

An overall performance metric was obtained by averaging the error values that resulted from applying the optimization technique to each project instance for each dataset. 33\% of instances are taken as test data in each dataset. This method provides a thorough assessment of the Firefly Algorithm's efficacy in improving weights for effort estimation by guaranteeing that the presented metrics represent the overall performance across all project instances in the dataset.

The following bar graphs and table provides an overview of the results:

\begin{figure}[h]
    \centering
    \includegraphics[width=0.7\textwidth, height = 180px]{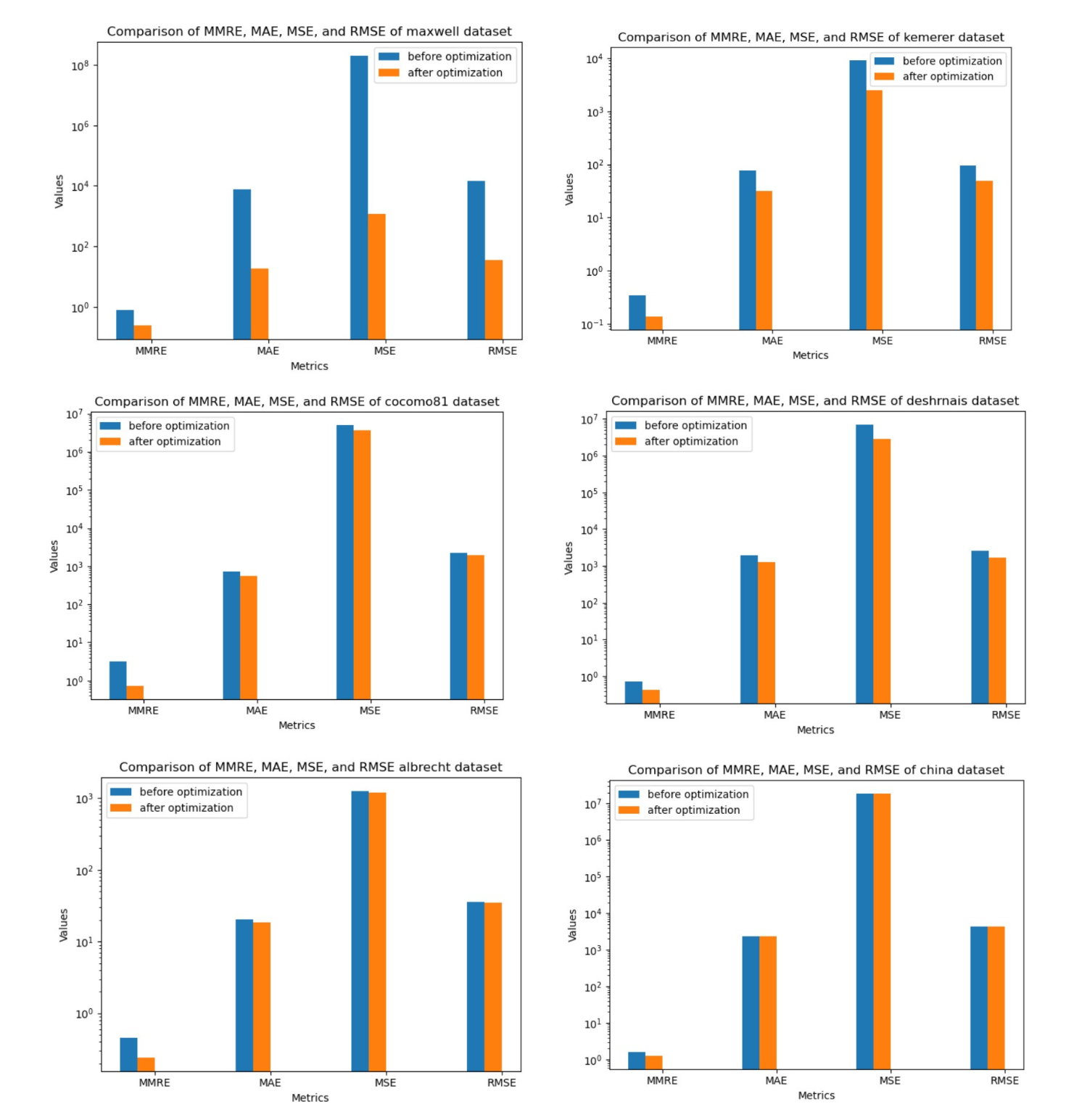}
    \caption{Comparison of performance metrics before and after optimization across six datasets.}
    \label{fig:results}
\end{figure}

\begin{table}[htbp]
\caption{Comparison of Estimation Accuracy (Before vs. After Optimization)}
\label{tab:experimental_results}
\centering
\resizebox{\columnwidth}{!}{%
\begin{tabular}{llcccc}
\toprule
\textbf{Dataset} & \textbf{Method} & \textbf{MMRE} & \textbf{MAE} & \textbf{MSE} & \textbf{RMSE} \\
\midrule
\multirow{2}{*}{\textbf{COCOMO81}} 
 & ABE & 3.2072 & 723.70 & 5.04e6 & 2233.3 \\
 & \textbf{FAABE} & \textbf{0.7188} & \textbf{62.24} & \textbf{3.63e6} & \textbf{1905.8} \\
\midrule
\multirow{2}{*}{\textbf{Desharnais}} 
 & ABE & 0.7264 & 1938.1 & 2.70e7 & 2629.5 \\
 & \textbf{FAABE} & \textbf{0.4270} & \textbf{1283.6} & \textbf{2.79e6} & \textbf{1670.2} \\
\midrule
\multirow{2}{*}{\textbf{China}} 
 & ABE & 1.5647 & 2346.6 & \textbf{1.87e7} & \textbf{4317.0} \\
 & \textbf{FAABE} & \textbf{1.2546} & \textbf{2332.0} & 1.89e7 & 4348.4 \\
\midrule
\multirow{2}{*}{\textbf{Albrecht}} 
 & ABE & 0.4573 & 20.41 & 1262.1 & 35.51 \\
 & \textbf{FAABE} & \textbf{0.2397} & \textbf{18.71} & \textbf{1194.7} & \textbf{34.56} \\
\midrule
\multirow{2}{*}{\textbf{Kemerer}} 
 & ABE & 0.3497 & 75.86 & 9102.5 & 94.31 \\
 & \textbf{FAABE} & \textbf{0.1371} & \textbf{31.42} & \textbf{2460.2} & \textbf{49.60} \\
\midrule
\multirow{2}{*}{\textbf{Maxwell}} 
 & ABE & 0.7859 & 7632.6 & 2.04e8 & 14266 \\
 & \textbf{FAABE} & \textbf{0.2397} & \textbf{18.71} & \textbf{1194.7} & \textbf{34.56} \\
\bottomrule
\end{tabular}%
}
\end{table}

\subsection{Performance Metrics}

\noindent \textbf{Mean Magnitude of Relative Error (MMRE):}
\[
\text{MMRE} = \frac{1}{k} \sum_{i=1}^{k} \left| \frac{Actual_i - Predicted_i}{Actual_i} \right|
\]
\noindent where:
\begin{itemize}
    \item \(k\) represents the total number of occurrences in the test set.
    \item \(Actual_i\) represents the actual effort for the \(i\)-th instance.
    \item \(Predicted_i\) represents the projected effort for the \(i\)-th instance.
\end{itemize}

\noindent \textbf{Mean Absolute Error (MAE):}
\[
\text{MAE} = \frac{1}{k} \sum_{i=1}^{k} \left| Actual_i - Predicted_i \right|
\]
\noindent where:
\begin{itemize}
   \item \(Actual_i\) represents the actual effort for the \(i\)-th instance.
   \item \(Predicted_i\) represents the projected effort for the \(i\)-th instance.
\end{itemize}

\noindent \textbf{Mean Squared Error (MSE):}
\[
\text{MSE} = \frac{1}{k} \sum_{i=1}^{k} (Actual_i - Predicted_i)^2
\]
\noindent where:
\begin{itemize}
   \item \(Actual_i\) represents the actual effort for the \(i\)-th instance.
   \item \(Predicted_i\) indicates the projected effort for the \(i\)-th instance.
\end{itemize}

\noindent \textbf{Root Mean Squared Error (RMSE):}
\[
\text{RMSE} = \sqrt{\frac{1}{k} \sum_{i=1}^{k} (Actual_i - Predicted_i)^2}
\]
\noindent where:
\begin{itemize}
    \item \(Actual_i\) represents the actual effort for the \(i\)-th instance.
    \item \(Predicted_i\) represents the projected effort for the \(i\)-th instance.
\end{itemize}

\section{Conclusion}
To optimize the weight parameters of the Analogy-Based Estimator (ABE) for effort estimate in software projects, the Firefly Algorithm (FA) was utilized in this work. Six datasets COCOMO81, Desharnais, Albrecht, Kemerer, China and Maxwell were used for the optimization. The outcomes show that the Firefly Algorithm's implementation significantly improved the performance indicators, such as MMRE, MAE, MSE, and RMSE. In particular, the optimization process decreased prediction errors, suggesting that FA improved the ABE model's accuracy and dependability for software development effort estimation. In future works, it is intended to propose and combine missing data imputation techniques with the
model in this study to strive for further improvement.

\section*{Acknowledgment}

The authors express sincere gratitude to Dr. Manjubala Bisi, Assistant Professor at NIT Warangal,  whose vision and guidance were the true inception of this work. The authors also thank Sarika Mustyala, research scholar in the Department of Computer Science and Engineering at NIT Warangal, for her mentoring and encouragement. Furthermore, the authors are deeply grateful to Dr. Tirimula Rao Benala, Assistant Professor at JNTU-GV College Of Engineering Vizianagaram, for his continual encouragement and support throughout this study.

\bibliographystyle{ieeetr}
\bibliography{references}

\end{document}